\documentclass[mypaper,8pt,twoside]{CoAst}
\usepackage{epsf,graphicx,fancyhdr}
\renewcommand{\chaptermark}[1]{\markboth{#1}{}}

\newcommand{\kopf}{\small\itshape Comm. in Asteroseismology \\ Contribution to the Proceedings of the Wroclaw HELAS Workshop, 2008}

\newcommand{\Authors}[1]{\begin{center}\normalsize\bf\sf #1 \end{center}}

\renewcommand{\author}[1]{\begin{center}\normalsize\bf\sf #1 \end{center}}
\newcommand{\Address}[1]{\begin{center}\small\sf #1 \end{center}}

\newcommand{\Session}[1]{{\vspace{3mm}\small \noindent  \hspace*{3mm} Session: } #1 \normalsize}

	\newcommand{\poster}{\small Poster \newline}

\newcommand{\References}[1]{\begin{flushleft}{\large References\\}\vspace*{2mm}\small #1 \end{flushleft}}

\newcommand{\chapterCoAst}[2]{\chapter[\sf\normalsize #1\\ \footnotesize \hspace*{5mm}by #2 \sf\normalsize][]{#1\\}\rhead[\fancyplain{}{\sf\footnotesize \center{#1}}]{\fancyplain{}{\sffamily\thepage}}\lhead[\fancyplain{\kopf}{\sffamily\thepage}]{\fancyplain{\kopf}{\sf\footnotesize \center{#2}}}}

\newcommand{\figureCoAst}[5]{\begin{figure}[#4]
\centering
\includegraphics*[#5]{#1}
\caption{#2}
\label{#3}
\end{figure}}

\renewcommand{\chaptername}{}

\def\rfr{\smallskip\par\noindent
        \hangindent=7truemm
        \hangafter=1}

\begin{document}
\sf

\chapterCoAst{Pulsating B and Be stars in the Magellanic Clouds}
{P.D. Diago, J. Guti\'{e}rrez-Soto, J.Fabregat, C. Martayan and J. Suso}
\Authors{P.D. Diago$^{1}$, J. Guti\'{e}rrez-Soto$^{1,2}$, J.Fabregat$^{1,2}$, C. Martayan$^{2,3}$ and J. Suso$^{1}$} 
\Address{
$^1$ Observatori Astron\`{o}mic de la Universitat de Val\`{e}ncia, Ed. Instituts d'Investigaci\'{o}, \\Pol\'{i}gon La Coma, 46980 Paterna, Val\`{e}ncia, Spain\\
$^2$ GEPI, Observatoire de Paris, CNRS, Universit\'e Paris Diderot, \\Place Jules Janssen 92195 Meudon Cedex, France\\
$^3$ Royal Observatory of Belgium, 3 Avenue Circulaire, B-1180 Brussels, Belgium
}

\noindent

\Session{ \poster } 


The $\kappa$-mechanism in $\beta$ Cephei and SPB stars has an important dependence on the abundance of iron-group elements, and hence the respective instability strips have a great dependence on the metallicity of the stellar environment. \textit{Pamyatnykh ~(1999)} showed that the $\beta$ Cephei and SPB instability strips practically vanish at $Z < 0.01$ and $Z < 0.006$, respectively.

The metallicity of the Magellanic Clouds (MC) has been measured to be around $Z=0.002$ for the Small Magellanic Cloud (SMC) and $Z=0.007$ for the Large Magellanic Cloud (LMC) (see \textit{Maeder et al.~1999} and references therein). Therefore, it is  expected to find a very low occurrence of $\beta$ Cephei or SPB pulsators in the LMC and no pulsator of this type in the SMC.

Our research has been focused in a sample of more than 150 stars for the LMC and more than 300 stars for the SMC (photometric time series had been provided by the MACHO project) 
for which \textit{Martayan et al.~(2006,~2007)} provided accurately determined fundamental astrophysical parameters. Our goal is to map the regions of pulsational instability  in the HR diagram for the low-metallicity environments as the MC.


The complete results of the analysis can be found in \textit{Diago et al.~(2008a)} for the SMC and \textit{Diago et al.~(2008b,~in preparation)} for the LMC. Many of the short-period variables have been found multiperiodic and some of them show the beating phenomenon due to the beat effect of close frequencies. In Table \ref{tabla} we resume the percentages of short-period variable stars compared with the results obtained by \textit{Guti\'{e}rrez-Soto et al.~(2007)} for the Milky Way (MW).

\begin{table}
\caption{Percentages of short-period variables in the MC and in the MW.}
\label{tabla}
\centering
\begin{tabular}{l c c c}
\hline\hline
			& MW	& LMC		& SMC		\\
\hline
Pulsating B stars	& 16\%	& 6.9\%		& 4.9\%		\\
Pulsating Be stars	& 74\%	& 14.8\%	& 24.6\%	\\
\hline
\end{tabular}
\end{table}

In the SMC, all pulsating B stars are restricted to a narrow range of temperatures (see Fig.~\ref{diagramaHR}). Moreover, all stars but one have periods longer than 0.5 days, characteristic of SPB stars. Thus, we suggest an observational SPB instability strip at the SMC metallicity shifted towards higher temperatures than in the Galaxy. We propose the hottest pulsating B star in our sample to be a $\beta$ Cephei variable. The reason is that it has two close periods in the range of the p-mode galactic pulsators, and it is the hottest pulsating star. If it is indeed a $\beta$ Cephei star, this would constitute an unexpected result, as the current stellar models do not predict p-mode pulsations at the SMC metallicities (see \textit{Miglio et al.~2007}).

For Be stars in the SMC, most of them are located inside or very close to this region, suggesting that they are g-mode SPB-like pulsators. Three stars are significantly outside the strip towards higher temperatures, all of them multiperiodic, with periods lower than 0.3 days. Therefore, we propose that these stars may be $\beta$ Cephei-like pulsators.

In the LMC the search for short period variables is more difficult than in the SMC because they show more outbursts and irregular variations that prevent us from carrying the frequency analysis. In spite of this difficulty, we have found 7 short-period variables among the B star sample and 4 among the Be star sample. As in the SMC, the hotter stars are those that are multiperiodic. Our work with the results of the LMC is ongoing and will be published by \textit{Diago et al.~(2008b,~in preparation)}.

\figureCoAst{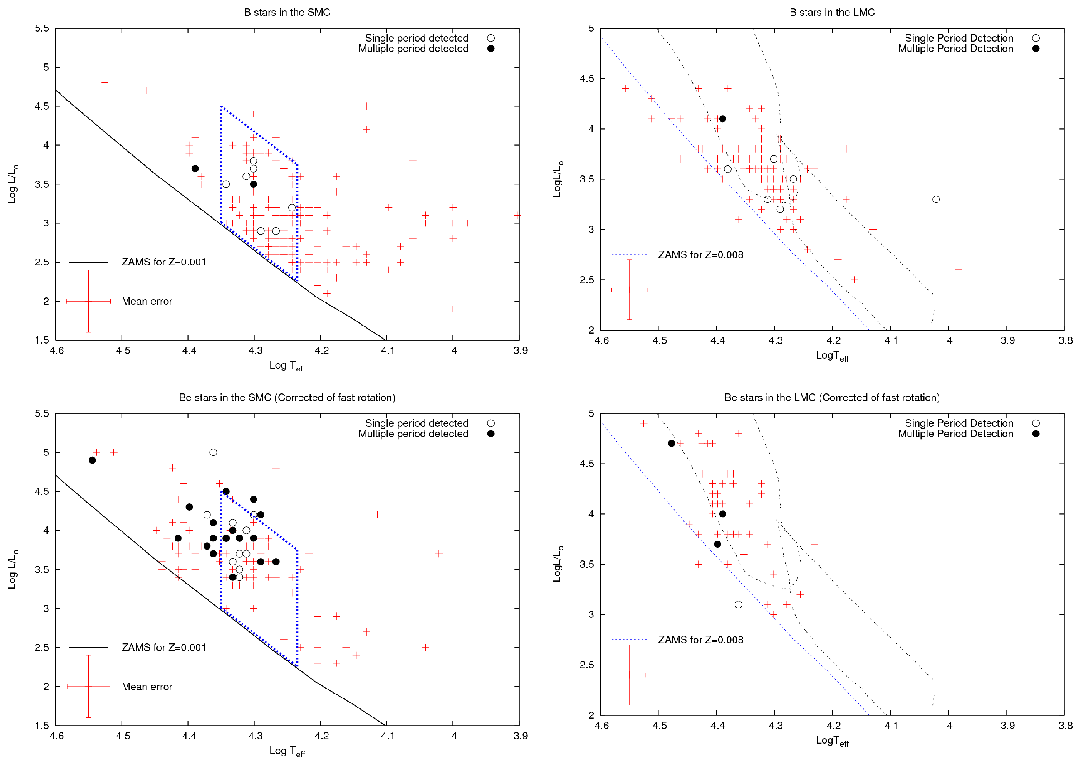}{Location of the B (top) and Be (bottom) star samples of the SMC (left pannels) and the LMC (right pannels) in the the HR diagram: single crosses represent stars in our sample, the empty circles represent single period detection and the filled ones multiple period detection. In the left panel (SMC), the dashed line delimits the suggested SPB instability strip for the SMC. In the right panel (LMC) the $\beta$ Cephei and the SPB boundaries at solar metallicities (\textit{Pamyatnykh 1999}) are plotted only for reference.}{diagramaHR}{h}{clip,angle=0,width=115mm}

\References{
\rfr Diago, P. D., Guti\'{e}rrez-Soto, J., Fabregat, J. \& Martayan, C. 2008, A\&A, 480, 179
\rfr Diago, P. D., Guti\'{e}rrez-Soto, J., Fabregat, J. \& Martayan, C. 2008, in preparation
\rfr Guti\'{e}rrez-Soto, J. , Fabregat, J., Suso, J., et al. 2007, A\&A, 476, 927
\rfr Maeder, A.,  Grebel, E. K., \& Mermilliod, J.-C. 1999, A\&A, 346, 459
\rfr Martayan, C., Floquet, M., Hubert, A.-M., et  al. 2006, A\&A, 452, 273
\rfr Martayan, C., Floquet, M., Hubert, A.-M., et  al. 2007, A\&A, 462, 683
\rfr Miglio, A., Montalb\'{a}n, J., \& Dupret, M.-A., et al. 2007, MNRAS, 375, L21
\rfr Pamyatnykh, A. A. 1999, Acta Astron., 49, 119
}

\end{document}